\documentclass[aps,prl,preprint,groupedaddress]{revtex4-2}
\usepackage{graphicx}  
\usepackage{dcolumn}   
\usepackage{bm}        
\usepackage{amssymb}   
\usepackage{amsmath}
\usepackage{url,hyperref}

\begin{document}

\title{Spectral properties of a mixed singlet-triplet Ising superconductor} 

\author{Sourabh Patil}
\affiliation{Fachbereich Physik, Universit\"{a}t Konstanz, D-78457 Konstanz, Germany}
\author{Gaomin Tang}
\affiliation{Department of Physics, University of Basel, Klingelbergstrasse 82, CH-4056 Basel, Switzerland}
\affiliation{Graduate School of China Academy of Engineering Physics, Beijing 100193, China  }
\author{Wolfgang Belzig}
\affiliation{Fachbereich Physik, Universit\"{a}t Konstanz, D-78457 Konstanz, Germany}
\email[]{wolfgang.belzig@uni-konstanz.de}

\date{\today}

\begin{abstract}
Conventional two-dimensional superconductivity is destroyed when the critical in-plane magnetic field exceeds the so-called Pauli limit. Some monolayer transition-metal dichalcogenides lack inversion symmetry and the strong spin-orbit coupling leads to a valley-dependent Zeeman-like spin splitting. 
The resulting spin-valley locking lifts the valley degeneracy and results in a strong enhancement of the in-plane critical magnetic field. In these systems, it was predicted that the density of states in an in-plane field exhibits distinct mirage gaps at finite energies of about the spin-orbit coupling strength, which arise from a coupling of the electron and hole bands at energy larger than the superconducting gap. In this study, we investigate the impact of a triplet pairing channel on the spectral properties, primarily the mirage gap and the superconducting gap, in the clean limit. Notably, in the presence of the triplet pairing channel, the mirage-gap width is reduced for the low magnetic fields. Furthermore, when the temperature is lower than the triplet critical temperature, the mirage gaps survive even in the strong-field limit due to the finite singlet and triplet order parameters. Our work provides insights into controlling and understanding the properties of spin-triplet Cooper pairs. 
\end{abstract}
\maketitle

\section{Introduction}
Superconductivity in the presence of magnetism has been a topic of great interest within the scientific community for several decades, and is responsible for many exotic properties.   
Usually, an external magnetic field destroys superconductivity by aligning the electron spins in the same direction. This is typical in conventional superconductors where Cooper pairs are formed by two electrons with opposite spins.  
The upper critical field is limited by the Pauli paramagnetic effect~\cite{Chandrasekhar,Clogston}. 
However, in some superconductors, the Pauli limit can be surpassed by forming Fulde-Ferrell-Larkin-Ovchinnikov (FFLO) states~\cite{matsuda} or by creating spin-triplet Cooper pairs so that the parallel-aligned spin configuration in Cooper pairs is not affected by Pauli paramagnetism~\cite{Aoki2001, huy, Aoki2012}. 
The spin-triplet Cooper pairings are important for spintronics 
applications~\cite{spintronics1} as they can generate and control spin currents. 
They are also proposed as a possible route to realize topological superconductivity which could be used to build robust quantum computers~\cite{Frolov20}. Therefore, understanding the properties of spin-triplet Cooper pairs is of great interest.

The transition-metal dichalcogenides serve as a platform for both exploring spin-triplet pairing physics and surpassing the Pauli limit in high-field superconductivity. It has been predicted that equal-spin triplet pairs can form in a superconducting few-layer 2H-NbSe$_2$ when an in-plane magnetic field is applied~\cite{Moeckli20}. Unlike conventional superconductors, monolayer transition metal dichalcogenides, such as NbSe$_2$, lack in-plane crystal inversion symmetry. This results in a spin-orbit interaction that generates an effective out-of-plane magnetic field, causing the electron spins to point out of the plane~\cite{MoS2_12,TMD_11}. Therefore, it is termed as Ising spin-orbit coupling (ISOC)~\cite{ Zhou16, Saito16, Lu15}, and the corresponding Ising superconductivity was experimentally found in numerous transition-metal dichalcogenides~\cite{Lu15, Saito16, Xi16, Xing17, Dvir18,Costanzo18, Lu18, delaBarrera18, Sohn18, Li20, Cho21, Hamill21, Idzuchi21, Ai21, Kang21, Kuz21}. ISOC is dependent on momentum and has opposite signs at the $K$ and $K'$ points of the hexagonal Brillouin zone. 
It prevents the spin directions from being realigned by an externally applied in-plane magnetic field, thus overcoming the Pauli limit to demonstrate high in-plane critical fields~\cite{Ilic17, Kuz21}.

Ising superconductors subjected to an in-plane magnetic field display unique features in their density of states, notably the emergence of additional half-gaps at finite energies of about the ISOC strength~\cite{mirage}. These newly discovered gaps, called mirage gaps, represent a mirroring of the main superconducting gap and are signatures of the equal-spin triplet finite-energy pairing correlations. Their width is determined by the interplay of the in-plane magnetic field and ISOC. 
In the previous work, the mirage gaps have only been studied in the context of the singlet pairing channel~\cite{mirage}. 
However, a recent experiment found that the superconducting gap in a few-layer NbSe$_2$ under an in-plane magnetic field was larger than predictions based solely on the singlet-pairing channel~\cite{Kuz21}. This discrepancy was attributed to the existence of a triplet-pairing channel, in which an equal-spin triplet order parameter couples with a singlet one to enhance the critical magnetic field. 
 
In this work, we investigate the mirage gaps of an Ising superconductor that consists of both singlet and triplet pairing channels~\cite{Kuz21}.
For a fixed temperature, the maximal mirage-gap width by varying the magnetic field decreases with increasing the critical temperature of the triplet pairing channel $T_{ct}$. 
When the magnetic field is extremely high and the temperature is lower than $T_{ct}$, the mirage-gap width remains finite due to the nonvanishing spin-singlet and spin-triplet order parameters. 
This contrasts with the case without considering the triplet-pairing channel where the mirage gaps always disappear at the critical field for the singlet order parameter~\cite{mirage}.

\section{Model and formalism}
An Ising superconductor with both a singlet order parameter $\Delta_s$ and an equal-spin triplet order parameter $\Delta_t$ can be described by a Bogoliubov–de Gennes Hamiltonian near the $\bm{K}$ ($\bm{K}'$) valley by neglecting the contribution from $\Gamma$ point~\cite{Kuz21}. 
By applying an in-plane magnetic field $\bm{B}$,
the effective Hamiltonian can be written in the Nambu basis ($c_{\bm{k}, \uparrow} , c_{\bm{k}, \downarrow} ,c^{\dagger}_{-\bm{k}, \uparrow} , c^{\dagger}_{-\bm{k}, \downarrow}  $) as
\begin{equation}
    H_{BdG} = \begin{bmatrix}
        H_0(\bm{k}) & \Delta i \sigma_y \\
        -\Delta i \sigma_y & -H_0^*(-\bm{k})
    \end{bmatrix} ,
\end{equation}
where $H_0(\bm{k})$ is given by
\begin{equation}
    H_0 (\bm{k} = \bm{p} + s\bm{K}) = \xi_p \sigma_0 + s\beta_{so} \sigma_z - \bm{B} \cdot \bm{\sigma} .
\end{equation}
Here, $s = +1$ corresponds to the valley $\bm{K}$ and $s = -1$ to $\bm{K}'$. The deviation of the momentum from $\bm{K}$ or $\bm{K}'$ is denoted by $\bm{p}$.
The Pauli matrices $\sigma_x$, $\sigma_y$, and $\sigma_z$ act on the spin space and $\sigma_0$ is the corresponding unit matrix.
The dispersion measured from the chemical potential is 
$\xi_p = p^2 /2m - \mu$. The ISOC strength is denoted by $\beta_{so}$. The Zeeman term arising from the in-plane magnetic field in the $x$-direction is $B_x \sigma_x$ which absorbs the factor of $g\mu_B /2$ containing the Land\'e $g$ factor and the Bohr magneton $\mu_B$. Note that the ISOC forces the spins to align out of the plane, whereas the in-plane magnetic field aims to align the spins within the plane. The superconducting order parameter can be written as \cite{Kuz21} 
\begin{equation}    
\Delta = \Delta_s +  i s \Delta_t \sigma_y .
\end{equation}
The eigenvalues of the Bogoliubov-de Gennes Hamiltonian are given by 
\begin{equation}
    \epsilon_{1(2)} = \sqrt{\xi_p^2 + \rho^2 + \Delta_s^2 + \Delta_t^2 \pm 2\sqrt{ \rho^2 \xi_p^2 + P^2}} 
\label{eigen}
\end{equation}
with $\rho = \sqrt{\beta_{so}^2 + B_{x}^2}$ and $P = B_x \Delta_s - \beta_{so} \Delta_t$. 
The position $\epsilon_0$ and the width $\delta$ of a mirage gap are obtained from the eigenvalues of the Hamiltonian, being given by
$\epsilon_0 = \pm (\epsilon_1 + \epsilon_2)/2$ and $\delta = \epsilon_1 - \epsilon_2$, respectively. At $\xi_p =0$ and in the limit of $\beta_{so} \gg \Delta_s, \Delta_t$, we have $\epsilon_0 \approx \rho$ and 
\begin{equation} \label{width}
    \delta \approx 2P / \rho  .
\end{equation}
The approximation indicates that the mirage-gap width is influenced by the interplay of the singlet and triplet order parameters.

Since the superconducting gap and the ISOC are much smaller compared to the Fermi energy, it allows us to use the formalism of quasiclassical Green's function~\cite{Eilenberger1968,LO1969,Belzig99,noneqSC,Eschrig15}. The structure of the Green's functions can be written as 
\begin{equation} \label{gg}
    \hat{g} (\hat{\bm{k}}, \epsilon) = 
    \begin{bmatrix}
    g_0 \sigma_0 + \bm{g} \cdot \bm{\sigma}  & 
    (f_0 \sigma_0 + \bm{f} \cdot \bm{\sigma}) i \sigma_y \\
    (\Bar{f_0} \sigma_0 + \Bar{\bm{f}} \cdot \bm{\sigma}^*) i \sigma_y  & \Bar{g}_0 \sigma_0 + \Bar{\bm{g}} \cdot \bm{\sigma}^* 
    \end{bmatrix}.
\end{equation}
The bar operation in the above expression is defined as $\hat{q} (\hat{\bm{k}}, \epsilon) = \hat{q} (-\hat{\bm{k}}, -\epsilon^*)^*$ with $q \in \{ g_0, f_0, \bm{g},\bm{f} \} $. 
For a homogeneous system in the clean limit, the Eilenberger equation can be written as  
\begin{equation} \label{Eilen}
    [\epsilon \sigma_0 \tau_3 - \hat{\Delta} - \hat{\nu} , \hat{g}]  = 0 .
\end{equation}
The order parameter term $\hat{\Delta}$ consisting of both the singlet and triplet components can be written as 
\begin{equation}
    \hat{\Delta} = 
    \begin{bmatrix}
    0 & (\Delta_s \sigma_0 + \bm{\Delta}_t \cdot \bm{\sigma}) i \sigma_y \\(\Bar{\Delta}_s \sigma_0 + \Bar{\bm{\Delta}}_t \cdot \bm{\sigma}^*) i \sigma_y & 0
    \end{bmatrix} ,
\end{equation}
with $\bm{\Delta}_t = (0, \; i s \Delta_t, 0)$. The Zeeman and ISOC fields are included in the term $\hat{\nu}$ as
\begin{equation}
    \hat{\nu} = s \beta_{so}\sigma_z \tau_3 - 
        \begin{bmatrix}
            \bm{B} \cdot \bm{\sigma} & 0 \\
            0 &  \bm{B} \cdot \bm{\sigma^*} 
        \end{bmatrix}.
\end{equation}
The Pauli matrices $\tau_1$, $\tau_2$, and $\tau_3$ act on the Nambu space and $\tau_0$ is the corresponding unit matrix.
By combining the Eilenberger equation with ${\rm tr}\big(\hat{g}\big)=0$ and the normalization condition $\hat{g}\hat{g}=\sigma_0\tau_0$, we can get the components of the Green's function. 
Defining $\Sigma= \Delta_{s}^2+\Delta_{t}^2-\epsilon^2+\rho^2$ and $\Omega = \sqrt{2\big(\Sigma - 2 \rho^2 + \sqrt{\Sigma^2-4P^2}\big)}$, we can express $g_0$, $f_0$, and $f_y$, respectively, as
\begin{align}
    g_0 &= 
    \frac{i \epsilon}{\Omega} \bigg( \frac{\Sigma}
    {\sqrt{\Sigma^2-4P^2}}+1 \bigg) , \\
    \label{eq_f0}
f_0 &= \frac{i}{\Omega} 
\bigg(\frac{\Delta_s \Sigma - 2 B_x P}{\sqrt{\Sigma^2 - 4P^2}} + \Delta_s \bigg) , \\ 
\label{eq_fy}
f_y &= \frac{1}{\Omega} 
\bigg(\frac{\Delta_t \Sigma + 2 \beta_{so} P}{\sqrt{\Sigma^2 - 4P^2}} + \Delta_t \bigg) .
\end{align}
We can use Green's functions $f_0$ in Eq.~\eqref{eq_f0} and $f_y$ in Eq.~\eqref{eq_fy} to derive the self-consistent equations for the singlet and the triplet order parameters. The cut-off frequencies $\Omega_{s(t)}$ are related to the critical temperatures $T_{c s(t)}$ and the coupling constants $\nu_{s(t)}$ by $\Omega_{s(t)} = 1.764 \; T_{c s(t)} \sinh{(1/\nu_{s(t)})}$. Assuming the same cut-off frequency for the singlet and the triplet pairing channels, we obtain $\nu_t = \nu_s / [1 + \nu_s \ln{(T_{cs}/ T_{ct})}]$. 
The singlet and triplet order parameters are coupled with the relations
\begin{equation}
\Delta_s = 2 \pi T |\nu_s| \sum_{\omega_n > 0} \frac{1}{\Omega} \bigg(\frac{\Delta_s \Sigma - 2 B_x P}{\sqrt{\Sigma^2 - 4P^2}} + \Delta_s \bigg) ,
\end{equation}
and 
\begin{equation}
\Delta_t = 2 \pi T |\nu_t| \sum_{\omega_n > 0} \frac{1}{\Omega} \bigg(\frac{\Delta_t \Sigma + 2 \beta_{so} P}{\sqrt{\Sigma^2 - 4P^2}} + \Delta_t \bigg) .
\end{equation}

At very high magnetic field and in the limit of $B_x \gg \beta_{so}, \epsilon, \Delta_s, \Delta_t$, one can show that
$\Delta_s = 2 \pi T |\nu_s| \sum_{\omega_n > 0} \beta_{so}\Delta_t / \big( B_x\sqrt{\omega_n^2 + \Delta_t^2} \big)$,
and $2 \pi T |\nu_t| \sum_{\omega_n > 0} \frac{1}{\sqrt{\omega_n^2 + \Delta_t^2}}=1$. This leads to $\Delta_s = \nu_s\beta_{so}\Delta_t / (\nu_t B_x)$ so that $P= (\nu_s / \nu_t -1) \beta_{so} \Delta_t$. Therefore, 
at high magnetic fields, the mirage gaps remain finite as long as the order parameters are finite.

\section{Numerical Results}
 
\begin{figure}[!ht] 
\begin{center}{}
\includegraphics[width=17cm]{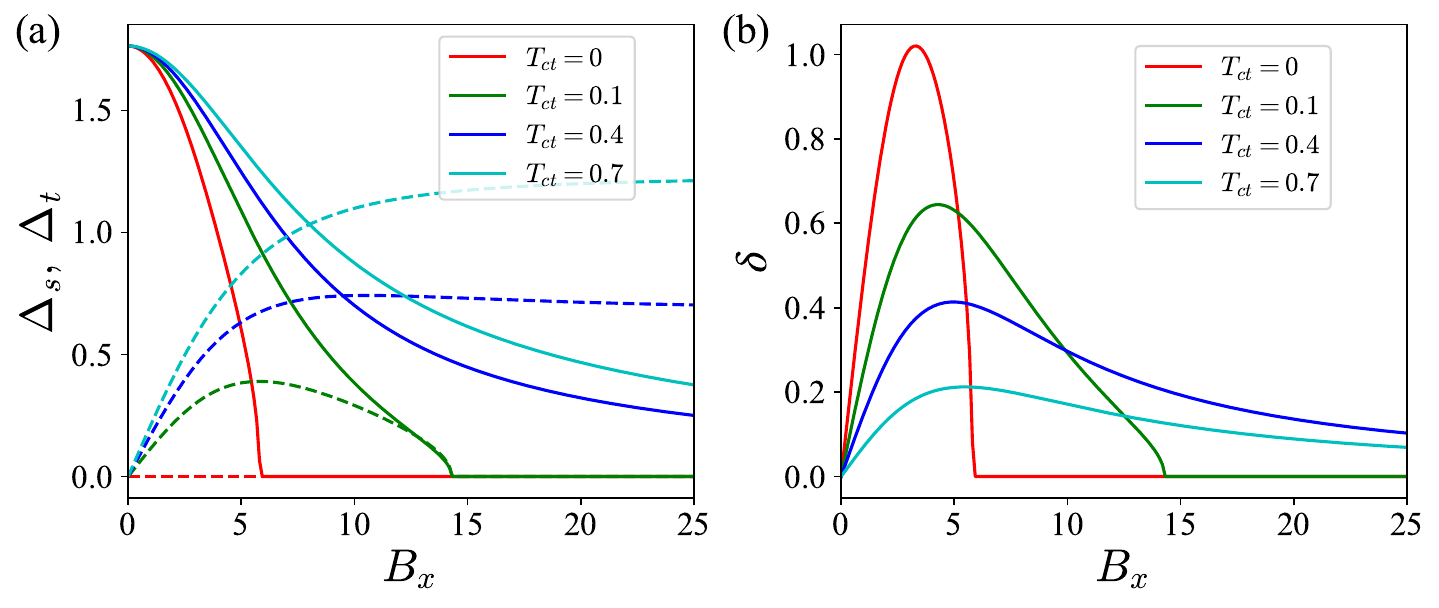}
\end{center}
\caption{(a) Singlet ($\Delta_s$, solid lines) and triplet ($\Delta_t$, dashed lines) order parameters versus in-plane magnetic field $B_x$ under different triplet critical temperatures $T_{ct}$. (b) The mirage-gap width $\delta$ versus $B_x$ under different $T_{ct}$. The temperature is $T=0.2$. All the quantities are in the units of $T_{cs}$.}
 \label{fig:1}
\end{figure}

In the numerical calculation, the ISOC strength is set as $\beta_{so} = 7 T_{cs}$. 
In Fig.~\ref{fig:1}(a), we plot the singlet ($\Delta_s$) and triplet ($\Delta_t$) order parameters versus the in-plane magnetic field $B_x$ by varying the triplet critical temperature $T_{ct}$ at a fixed temperature $T$.
The behavior of the order parameters in the case of $T > T_{ct}$ and $T<T_{ct}$ will be discussed separately in the following. 
For the case where $T > T_{ct}$, with an increasing magnetic field, the singlet order parameter $\Delta_s$ decreases and finally vanishes at the critical field.
This is attributed to the pair-breaking effect of the magnetic field which tries to align the spins of the spin-singlet Cooper pairs in its direction. 
On the other hand, the in-plane magnetic field induces the equal-spin triplet pairings which are coupled to the singlet pairings.
The triplet order parameter $\Delta_t$ first increases with increasing the in-plane magnetic field since the field attempts to align the spins in the plane which is favorable for the formation of triplet pairs. 
It then decreases due to its coupling with $\Delta_s$ as seen from the self-consistent gap equations. Finally, both $\Delta_s$ and $\Delta_t$ vanish at the same critical magnetic field for $T>T_{ct}$. 
For the case where $T < T_{ct}$, at small magnetic fields, $\Delta_s$ decreases, and $\Delta_t$ begins to increase similar to the case where $T > T_{ct}$. The behavior of both the order parameters differ from that of $T > T_{ct}$ case at high magnetic fields. Both $\Delta_s$ and $\Delta_t$ reach saturated values in the high magnetic field limit, instead of vanishing at a critical field. This is because the triplet order parameter is preserved by the in-plane field as the field favors the formation of triplet pairs.
Consequently, $\Delta_s$ does not vanish due to its coupling with $\Delta_t$.

Figure.~\ref{fig:1}(b) shows the mirage-gap width $\delta$ for different triplet critical temperatures $T_{ct}$.
The maximal mirage-gap width by varying the magnetic field decreases with increasing $T_{ct}$.
The mirage gaps vanish at the critical in-plane field when $T> T_{ct}$. 
This is due to the dependence of mirage-gap width on the order parameters which vanish at the critical field.  
On the other hand, the mirage gaps are finite even at extremely high in-plane fields in the scenarios where $T < T_{ct}$. 
This is because the order parameters remain finite in the high field limit at $T < T_{ct}$.

\section{Conclusion}
To conclude, we have studied the effect of a triplet pairing channel on the spectral properties of an Ising superconductor.
The presence of a triplet order parameter reduces the maximal mirage-gap width by varying the in-plane magnetic field. 
Notably, at a temperature lower than $T_{ct}$, both the order parameters and the mirage gap are finite even at very high fields. In contrast, for the case where the temperature is higher than $T_{ct}$, both the order parameters and the mirage gaps vanish at the critical field. This work provides better understanding about the spin-triplet pairings in Ising superconductors.

\section{Conflict of Interest Statement}
The authors declare that the research was conducted in the absence of any commercial or financial relationships that could be construed as a potential conflict of interest.

\section{Author Contributions}
SP, GT, and WB contributed to conception of the study. SP and GT performed analytical and numerical calculations. SP wrote the first draft of the manuscript. All authors contributed to manuscript revision, read, and approved the submitted version.

\section{Funding}
We acknowledge funding by the Deutsche Forschungsgemeinschaft (DFG, German Research Foundation)–Project-ID 443404566 - SPP 2244.

\section{Acknowledgments}
We acknowledge useful discussions with M. Aprili, C. Bruder, A. Di Bernardo and D. Nikoli\'{c}.

\bibliography{refs}  

\end{document}